\documentclass{article}
\makeatletter

\@addtoreset{equation}{section}
\makeatother
\newtheorem{theo}{Theorem}[section]
\newtheorem{rema}[theo]{Remark}
\newtheorem{lemm}[theo]{Lemma}
\newtheorem{propo}[theo]{Proposition}

\newtheorem{defi}[theo]{Definition}
\newtheorem{ex}[theo]{Example}
\newcommand{\m}{{\cal M}}   
\newcommand{\mo}{{\cal M}_0}  
\newcommand{\be}{\begin{equation}}
\newcommand{\ee}{\end{equation}}

\newcommand{\cvd}{{\rule{0.5em}{0.5em}}\smallskip}

\font\ddpp=msbm10 
\def\R{\hbox{\ddpp R}}    
\def\N{\hbox{\ddpp N}}    
\def\Z{\hbox{\ddpp Z}}

\title{{\bf Causality and Conjugate Points in General Plane Waves}}

\author{J.L. FLORES,
M. S\'ANCHEZ \\
\\
{\small Departamento de Geometr\'{\i}a y Topolog\'{\i}a}\\
{\small Facultad de Ciencias, Universidad de Granada}\\
{\small Avenida Fuentenueva s/n,
18071 Granada, Spain}\thanks{The authors acknowledge some comments by Dr. C. Hillman. Partially supported by a MCyT-FEDER Grant BFM2001-2871-C04-01. The first-named author is also supported by a MECyD Grant EX-2002-0612}\\
{\small jflores@ugr.es, sanchezm@ugr.es}\\
}

\date{}

\begin{document}
\textwidth=140mm
\textheight=200mm
\parindent=5mm
\maketitle

\begin{center} {\small \bf Abstract.} \end{center}
{\small

\noindent
Let $\m = \mo \times \R^2$ be a pp--wave type spacetime  endowed with the metric
$\langle\cdot,\cdot\rangle_z = \langle\cdot,\cdot\rangle_x
+ 2\ du\ dv + H(x,u)\ du^2$, where $(\mo, \langle\cdot,\cdot\rangle_x) $ is any Riemannian manifold and $H(x,u)$ an arbitrary function. We show that the  behaviour of $H(x,u)$ at spatial infinity determines the causality of $\m$, say: (a) if $-H(x,u)$ behaves subquadratically  (i.e, essentially  $-H(x,u)  \leq  R_1(u) |x|^{2-\epsilon} $ for some  $\epsilon >0$ and large distance  $|x|$ to a fixed point)
and the spatial part $(\mo, \langle\cdot,\cdot\rangle_x) $ is complete,
then the spacetime $\m$ is globally hyperbolic, (b) if $-H(x,u)$  grows at most quadratically  (i.e, $-H(x,u)  \leq  R_1(u) |x|^{2}$ for large $|x|$) then it is strongly causal and (c) $\m$ is always causal, but there are non-distinguishing examples (and thus, non-strongly causal), even when $-H(x,u)  \leq  R_1(u) |x|^{2+\epsilon} $, for small $\epsilon >0$.

Therefore, the classical model $\mo = \R^2$, $H(x,u) = \sum_{i,j} h_{ij}(u) x_i x_j
(\not\equiv 0)$, which is known to be strongly causal but not globally hyperbolic, lies in the
critical quadratic situation with complete $\mo$. This must be taken into account for realistic applications. In
 fact, we argue that $-H$ will be subquadratic (and the spacetime globally hyperbolic) if $\m $ is asymptotically flat. The relation of these results with the notion of astigmatic conjugacy and the existence of conjugate points is also discussed.
}\\
\\
{\small\sl Keywords:}
{\small Gravitational waves, plane fronted waves, causality, global hyperbolicity,
conjugate points, astigmatic conjugacy.}\\
\\
{\small\sl PACS:} {\small 04.20.Gz, 04.20.Jb.}\\
{\small\sl 2000 MSC:} {\small 53C50, 83C35.}


\newpage

\section{Introduction}
Classically, plane fronted waves are studied in General Relativity by means of the
model $(\R^4,ds^2)$
\be
\label{planefrontedmetric}
d s^2 = dx_1^2 + dx_2^2 + 2\ du\ dv + H(x_1, x_2,u)\ du^2 ,
\ee
$(x_1, x_2,v,u) \in \R^4$, where the (non identically null) function $H : \R^3 \to \R$
can be written as
\begin{equation} \label{quadratic}
H(x,u) = \sum_{i,j=1}^2 h_{ij}(u) x_i x_j
\end{equation}
for some symmetric functions $h_{ij}$.
Particular cases are {\em (plane symmetric) electromagnetic waves} ($h_{ij}(u) = h(u)\delta_{ij}$) and  {\em  (gravitational) plane waves} ($\sum_i h_{ii} \equiv 0$). In spite of their interest, some idealizations of these models (as the infinite transversality of the wave or the identically null curvature of $(\R^2, dx^2_1+ dx_2^2)$)  may imply unrealistic predictions, especially from a global viewpoint. In fact, some authors have studied finite models of plane waves as well as extensions of the exact model (\ref{planefrontedmetric}) to different situations (see, for example, 
Refs. \cite{Yu2}---\cite{Bi}). 

We will focus on one of the outstanding  global properties of a spacetime,  its causal structure and, in particular, its possible global hyperbolicity. In fact, this property affects not only to its geometry or to the possibility to specify Cauchy data for Einstein equations, but also to quantization \cite{Yu5}, \cite{LS}. Penrose \cite{Pe} proved that, in the model
(\ref{planefrontedmetric}), (\ref{quadratic}), there exists a  sequence of null geodesics which converges on a pair of nonintersecting null geodesics. This focusing property has remained as one of the folk characteristics of plane waves and, as Penrose himself proved, it  forbides not only global hyperbolicity but also the possibility of global embeddings in the flat space
$\R^N$ with arbitrary signature.
In a series of articles, Ehrlich and Emch \cite{EE1}, \cite{EE2}, \cite{EE3} studied  systematically
the focusing property, by introducing the concept of {\em astigmatic conjugacy} for pairs of values of the variable $u$. Moreover, they determined the precise causal hierarchy of exact gravitational waves, by showing that they are
causally continuous (and thus, strongly causal) but not causally simple (neither globally hyperbolic), see also the book \cite{BEE}. Other causality properties of waves type
(\ref{planefrontedmetric}) can be seen in \cite{HuRa}.

The authors, in collaboration with A.M. Candela, have introduced the following generalization of the exact model \cite{CFSa}. A  {\em (general) plane fronted wave} (PFW)\footnote{We will use the name PFW in what follows, in agreement to \cite{CFSa} and with no further pretension. But there is some possibility of confusion with the related names in the literature. Some widely spread names for the gravitational (i.e. vacuum) case are \cite[Ch. 8]{Bi}: (a) plane fronted wave: vacuum spacetime admitting a shearfree geodesics null--congruence and an orthogonal distribution of spacelike 2--surfaces, (b) pp-wave: vacuum metric satisfying (\ref{planefrontedmetric}), and (c) plane wave: pp-wave satisfying (\ref{quadratic}) (the name ``plane wave'' is then more restrictive than ``plane fronted wave''). At any case,
our models include all (vacuum or not) pp-waves and, then, an alternative  name may be GppW ---general pp-wave.}  is a product manifold
$\m = \mo \times \R^2$ endowed with the metric
\be
\label{metric}
\langle\cdot,\cdot\rangle_z = \langle\cdot,\cdot\rangle_x
+ 2\ du\ dv + H(x,u)\ du^2 ,
\ee
where $(\mo, \langle\cdot,\cdot\rangle_x)$ is an arbitrary (connected) Riemannian manifold,
the variables $(v,u)$ are the natural coordinates
of $\R^2$ and the smooth scalar field $H(x,u)$ on $\mo\times \R$ is also arbitrary
(the subscripts $z $ and $x$ of the metric will be suppressed in what follows, if there is no possibility of  confusion). In \cite{CFSa} it is shown that some global properties of the geodesics of a PFW (geodesic connectedness, completeness) depend on the behaviour of $H(x,u)$ at spatial infinity, i.e., when the distance of $x$ to a fixed point $\bar x$ becomes arbitrarily large. Moreover, a quadratic behaviour such as (\ref{quadratic}) becomes critical, in the sense that small perturbations either in the superquadratic or in the subquadratic direction may introduce significative qualitative differences. The main aim of the present paper is to show that this also holds for the causal structure and, in fact, perturbations in the subquadratic direction for $-H$ will yield global hyperbolicity, while in the superquadratic direction may destroy strong causality or even to be distinguishing.


More precisely, we will say that $-H(x,u)$ {\em behaves subquadratically at spatial infinity} if
there exist $\bar x \in \mo$ and continuous functions $R_1(u), R_2(u) (\geq 0)$, $p(u) < 2$ such that:
 \begin{equation} \label{subquadratic}
-H(x,u) \le R_1(u) d^{p(u)}(x,\bar x) + R_2(u) \quad
\forall (x,u) \in \mo \times \R,
\end{equation}
where $d$ is the distance canonically associated to the Riemannian metric on $\mo$. Less restrictively, if (\ref{subquadratic}) holds with $p(u) \equiv 2$ then
$-H(x,u)$ behaves {\em (at most) quadratically at spatial infinity}.
Then, we will prove:
\begin{itemize}
\item (Section \ref{s2}.) If no restriction on $H(x,u)$ is imposed, the PFW
is causal, but not necessarily distinguishing (therefore, neither
strongly causal). In fact, a general reasoning shows that PFW's
are non-distinguishing when $-H$, in addition to
superquadratic (in the sense of Proposition \ref{remnondis}), is non-negative, and ${\cal
M}_0$ is complete. As consequence,
simple non-distinguishing counterexamples can be found even under
the additional assumption $p(u) < 2+ \epsilon$, for (small)
$\epsilon>0$, Example \ref{ex1}.
\item (Section \ref{s3}.) If $-H(x,u)$ behaves 
at most quadratically at spatial infinity, then the PFW is
strongly causal, Theorem \ref{tsc}. So, this property of plane
waves (\ref{quadratic}) is general for all other PFW's with the
same asymptotic behaviour.

\item (Section \ref{s4}.) If $-H(x,u)$
behaves   subquadratically at spatial infinity and the
Riemannian distance $d$ on $\mo$ is  complete, then the
spacetime is not only strongly causal but also globally
hyperbolic, Theorem \ref{tgh}. The proof of global hyperbolicity
can be also used to give particular
examples of quadratic and superquadratic PFW's which are also
globally hyperbolic (Example \ref{exglobhyp}).
\end{itemize}
The importance of these results becomes apparent, because (apart from problems of quantization, where the asymptotic behaviour is fundamental) the classical model
(\ref{planefrontedmetric}), (\ref{quadratic}) lies exactly in the limit quadratic case, with $\mo$ complete. In fact, Penrose \cite{Pe} also argued that, in order to be physically meaningful, the exact model must be modified in some sense to obtain asymptotic flatness. This suggests that physically meaningful models for plane waves will be globally hyperbolic. In Section \ref{s45} we explore this possibility in two steps (see the conclusions in Remark \ref{recont}): (a) the energy conditions are  characterized for PFW's
(Proposition \ref{45a}), showing explicit examples with $-H(x,u)$ sub and superquadratic, and (b) we argue that asymptotic flatness should imply the  vanishing of the eigenvalues of Hess$_xH$ at infinity, and prove that, for complete ${\cal M}_0$, this condition implies $-H(x,u)$ is subquadratic (Proposition \ref{45b}), i.e., global hyperbolicity.

On the other hand, as the focusing  of null geodesics is an essential property
for  plane waves, conjugate points of a general PFW are also
studied (Section \ref{s5}). Basically, we show that the conjugate
points along any geodesic are equal to the conjugate points of the
trajectories for a (positive-definite) Riemannian problem of a
particle under a potential, Proposition \ref{reeff}. This allows
to control the existence of conjugate points in a precise way, and
may suggest their existence in some cases, even though the
focusing property of the classical model
(\ref{planefrontedmetric}), (\ref{quadratic}) cannot be expected
in general.


\section{PFW's are not necessarily distinguishing}\label{s2}

In what follows, the signature of spacetimes is chosen $(-,+,
\dots , +)$, and a tangent vector $w$ will be timelike (resp.
lightlike, causal)  if $\langle w, w\rangle <0$ (resp., $\langle
w, w\rangle =0$ and $w\neq 0$;  $w$ is  either lightlike or
timelike); vector 0 is spacelike. ${\cal M}={\cal
M}_{0}\times\R^{2}$ will be equipped with the metric
(\ref{metric}), and we will assume differentiability $C^2$ as a
simplification ($C^1$ would be enough for most purposes). We will
fix the time-orientation such that the lightlike vector field
$\partial_v$,  is past directed; thus, the lightlike vector field
$\partial_u -\frac{1}{2} H\partial_v$ will be future-directed.
Easily, $\partial_v = \nabla u$ is a parallel vector field and,
for any future-directed causal curve $z(s)=(x(s),v(s),u(s))$, \be
\label{udot} \dot u(s) = \langle \dot z(s), \partial_v\rangle \geq
0, \ee with strict inequality  if $z(s)$ is timelike.

From (\ref{udot}),
 PFW's cannot contain closed timelike curves (they are {\em chronological}). Moreover,
 $u$ is a {\em quasi-time} function (i.e., a function $f$ such that: (a) $\nabla f$ is everywhere causal and past directed, and (b) every lightlike geodesic $\gamma$ with $f\circ \gamma$ constant is injective) and, as a consequence, PFW's are {\em causal}, i.e., can neither contain closed causal curves (see, for example, \cite[Scholium 4.11]{EE1}). Nevertheless, PFW's are not necessarily strongly causal, as Proposition \ref{remnondis} shows. This result provides examples of PFW's, even as in (\ref{planefrontedmetric}), which are not distinguishing. Recall
that {\em distinguishing} is the causality condition strictly
between causal and strongly causal (see \cite[p. 73]{BEE}), and it
means the equivalence among the three following  conditions: (i)
$I^{+}(P)=I^{+}(Q)$; (ii) $I^{-}(P)=I^{-}(Q)$; and (iii) $P=Q$.

\begin{propo}\label{remnondis}  A PFW is
non-distinguishing if ${\cal M}_{0}$ is complete, $H$ is
non-positive ($H\leq 0$) and $-H$ is superquadratic in the following sense:
there exists a sequence $\{y_{n}\}_{n}\subseteq {\cal M}_{0}$ with
$d(y_{n},\bar{x})\rightarrow\infty$ such that
\begin{equation}\label{ya}
-H(y_{n},u)\geq R_{1}\cdot
d^{2+\epsilon}(y_{n},\bar{x})+R_{2}\quad\;\;\hbox{for all}\;\;
u\in\R,
\end{equation}
for some $\bar{x}\in {\cal M}_{0}$ and $\epsilon, R_{1}, R_{2}\in\R$
with $\epsilon,R_{1} >0$.
\end{propo}
{\it Proof.} Fix $z_{0}=(x_{0},v_{0},u_{0})$. In order to check that the implication (i) $\Rightarrow$ (iii) in the definition of distinguishing above fails, it is enough to show
that $I^+(z_0)= \{(x, v, u): x\in {\cal M}_{0}, v\in \R,
u_0<u\}={\cal M}_{0}\times \R\times (u_{0},\infty),$ for all
$x_{0}$, $v_0$.

Recall that, for any future-directed timelike
curve, $\dot u(s)
>0$ (see (\ref{udot})) and, thus, $I^+(z_0)\subseteq {\cal
M}_{0}\times \R \times (u_0, \infty)$. In the remainder we will
show the converse, that is, fixed $z_1= (x_1,v_1,u_1)$ with
$u_0<u_1$, our aim is to construct  a piecewise smooth
future-directed timelike curve $z(s)=(x(s),v(s),u_0+s \Delta u)$,
$\Delta u = u_1-u_0$, $s \in [0,1]$ from $z_0$ to $z_1$.

Define for $0<\delta <1$ and any natural number $n\in \N$,
the piecewise smooth curve
\begin{equation}\label{yaa}
x_n(s)=\left \{\begin{array}{ll} \alpha_{n}(s) &
\quad\;\;\hbox{if}\;\; s\in [0,\frac{1}{2}-\frac{\delta}{2}] \\
y_{n} & \quad\;\;\hbox{if}\;\; s\in
[\frac{1}{2}-\frac{\delta}{2},\frac{1}{2}+\frac{\delta}{2}] \\
\beta_{n}(s) & \quad\;\;\hbox{if}\;\; s\in
[\frac{1}{2}+\frac{\delta}{2},1],\end{array}\right.
\end{equation}
where $\alpha_{n}:[0,\frac{1}{2}-\frac{\delta}{2}]\rightarrow
{\cal M}_{0}$,
$\beta_{n}:[\frac{1}{2}+\frac{\delta}{2},1]\rightarrow {\cal
M}_{0}$ are minimizing geodesics for the Riemannian distance on $\mo$
joining $x_{0}$ with $y_{n}$ and
$y_{n}$ with $x_{1}$, respectively.

Now, for any arbitrary function $v_n(s)$,
the piecewise smooth curve  $z_n(s)=(x_n(s),v_n(s),u_0+s\Delta u)$ joins $z_0$ with $(x_{1}, v, u_1)$,
for some (uncontrolled) $v$. Taking into account the
expression of the metric (\ref{metric}), define $v_n(s)$ to obtain a
constant-speed timelike curve. That is, for some $E<0$ put:
\begin{equation}\label{virgi}
v_n(s)-v_0 = \frac{1}{2\Delta u}\int_{0}^{s}(E-\langle
\dot{x}_n(\sigma),\dot{x}_n(\sigma)\rangle -(\Delta u)^2
H(x_n(\sigma),\sigma)) d\sigma.
\end{equation}
(Even though $z_n(s)$ is only piecewise smooth, it satisfies
$\langle\partial_{v},\dot{z}_{n}(s)\rangle =\Delta u >0$, and
thus, it is future-directed). Therefore, using (\ref{yaa}),
(\ref{ya}) and the fact that $H(x,u)\leq 0$, we obtain

$$
v_n(1)-v_{0}\geq \frac{E}{2\Delta u}
-\frac{1}{2\Delta u}\left(
\int_{0}^{\frac{1}{2}- \frac{\delta}{2}}
\langle \dot{x}_n(\sigma),\dot{x}_n(\sigma)\rangle d\sigma
+
\int_{\frac{1}{2}+\frac{\delta}{2}}^1
\langle \dot{x}_n(\sigma),\dot{x}_n(\sigma)\rangle d\sigma \right) $$
$$-
\frac{\Delta u}{2}
\int_{\frac{1}{2}-\frac{\delta}{2}}^{\frac{1}{2}+\frac{\delta}{2}}
H(y_n,\sigma ) d\sigma $$
\begin{equation}\label{virgii}
\ge \frac{E}{2\Delta u}-\frac{d^{2}(y_{n},x_{0})+d^{2}(y_{n},x_{1})}{(1-\delta)\Delta
u}+\frac{\Delta u}{2} \delta (R_{1}\cdot
d^{2+\epsilon}(y_{n},\bar{x})+R_{2}).
\end{equation}
From (\ref{virgi}), $v_n(1)$ goes to $-\infty$ when $E$ goes to
$-\infty$ and $n\in \N$ is fixed. On the other hand, from
(\ref{virgii}) $v_n(1)$ goes to $+\infty$ when $n$ goes to $+\infty$
and $E\in (-\infty,0)$ is fixed\footnote{This is the exact point where $R_1, \epsilon>0$ is needed.}. Therefore, for $n$ big enough and
varying $E\in (-\infty,0)$, the value of $v_n(1)$ can reach $v_{1}$,
as required. \cvd
\begin{ex}\label{ex1} {\rm
An obvious example included in Proposition \ref{remnondis} is the following: ${\cal M}_{0}=\R^{2}$ and $H(x,u)$ equal to
$-|x|^{2+\epsilon}$, $\epsilon >0$ (if required, $H$ can be modified around $x=0$ in order to obtain a smooth function).
}\end{ex}
This wide family of non-distinguishing examples shows a clear
difference with classical plane waves (\ref{planefrontedmetric}), (\ref{quadratic}), which are always strongly
causal. In fact, the counterexamples are possible because $-H$
behaves at spatial infinity faster than $|x|^{2}$; for a
 behaviour at most as $|x|^2$ strong causality is ensured, as proven in the next section.  But, of course,
a superquadratic  behaviour does not imply necessarily such a bad
causal behaviour: globally hyperbolic examples both, quadratic and
superquadratic, will be constructed in Example \ref{exglobhyp}.

On the other hand, it is not difficult to check that the arguments for the classical case
(\ref{planefrontedmetric}), (\ref{quadratic}) apply, showing that
any PFW is {\em $u$--causally convex}
(according to \cite[Definition 4.3]{EE1}); moreover,  if  $(\mo,\langle\cdot,\cdot\rangle_{x})$
has no geodesic loops (i.e., if any geodesic $x(s)$ of $\mo$ with $x(0)=x(1)$ is necessarily constant) then the PFW is also
{\em causally disconnected by a compact subset} (see the proof of \cite[Proposition 4.15]{EE1}).

\section{Sufficient hypothesis for Strong Causality} \label{s3}

 Next, we will see how the quadratic behaviour of   $-H(x,u)$ at spatial infinity is enough to ensure strong causality, i.e.:
\begin{theo} \label{tsc}
Any PFW (according to (\ref{metric})) such that
 $-H(x,u)$ grows at most quadratically at infinity is strongly causal, for any (complete or not) ${\cal M}_0$.
\end{theo}
In order to prove strong causality (and, when necessary, global hiperbolicity), causal curves with endpoints in a controlled subset must be also controlled between the endpoints. To this aim  the inequality in Lemma
\ref{lsc2} below will be used systematically. But, in order to prove this inequality, first we will need the following
technical one. Essentially, this means that, for curves $x(u)$ defined on an interval $[u_0,u_0+\epsilon]$ in the Riemannian manifold $\mo$, the smaller $\epsilon >0$ we choose, the bigger integral of the energy $|\dot x|^2$ (in comparison with the integral of the
square distance of $x(s)$) we get.

\begin{lemm} \label{lsc}
Fix $\epsilon >0$.
Then, for any piecewise smooth curve $x:[u_{0},u_{1}]\rightarrow
{\cal M}_{0}$ with $0< u_1-u_0 < \epsilon$:
\begin{equation}\label{toto}
\int_{u_{0}}^{u}\langle \dot{x}(s),\dot{x}(s)\rangle ds\geq
\frac{1}{\epsilon^{2}} \int_{u_{0}}^{u} d^{2}(x(s),x(u_{0}))
ds.
\end{equation}
\end{lemm}
{\bf Proof.} Clearly, for all $u\in [u_{0},u_1]$,
\[
d(x(u),x(u_{0}))\leq
\int_{u_{0}}^{u}\sqrt{\langle\dot{x}(s),\dot{x}(s)\rangle} ds\leq
\sqrt{u-u_{0}}\cdot \sqrt{\int_{u_{0}}^{u}\langle
\dot{x}(s),\dot{x}(s)\rangle ds}
\]
(last inequality by Cauchy-Schwarz). Therefore,
\[
d^{2}(x(u),x(u_{0}))\leq
(u-u_{0})\int_{u_{0}}^{u}\langle\dot{x}(s),\dot{x}(s)\rangle ds
\]
and, if $u_{0}\leq s\leq u$, then
\begin{equation}\label{fgh}
d^{2}(x(s),x(u_{0}))\leq
(s-u_{0})\int_{u_{0}}^{s}\langle\dot{x}(\bar s),\dot{x}(\bar s)
\rangle d\bar s\leq (u-u_{0})\int_{u_{0}}^{u}\langle\dot{x}(\bar
s),\dot{x}(\bar s)\rangle d\bar s.
\end{equation}
Finally, by integrating in $s$ the extreme terms of (\ref{fgh}),
we have
\[
\int_{u_{0}}^{u}d^{2}(x(s),x(u_{0}))ds
\leq (u-u_{0})^{2}\int_{u_{0}}^{u}\langle
\dot{x}(s),\dot{x}(s)\rangle ds \leq
\epsilon^{2}\int_{u_{0}}^{u}\langle \dot{x}(s),\dot{x}(s)\rangle
ds,
\]
and (\ref{toto}) is obtained. \cvd
\smallskip

\noindent In the following crucial lemma, for causal curves
$\alpha(u) = (x(u), v(u), u)$, with domain
$[u_{0},u_{0}+\epsilon]$ and fixed $x(u_{0})=x_{0}$, the integral
of the energy of the component $x(u)$ is  upper bounded in
terms of the extremes of the components $(v(u), u)$ of $\alpha$,
with a bound {\em independent of the chosen curve}.
\begin{lemm} \label{lsc2}
Assume that $-H(x,u)$ behaves quadratically (inequality (\ref{subquadratic}) holds with
$p(u)\equiv 2$), and
fix $u_{0}\in\R$ and  a bounded subset $B \subset {\cal M}_{0}$. There exist $\epsilon>0$
and $R_2'>0$ such that, for any $u$-reparametrized causal curve
$\alpha(u) = (x(u), v(u), u)$, $u \in [u_0, u_1]$ with $u_1-u_0
\leq \epsilon$ and $x(u_0)=x_0 \in B$, one has:
\be \label{aux3} v(u)
-v(u_0) - R_2'(u-u_0) < -\frac{1}{4} \int_{u_0}^u \langle
\dot{x}(\bar u),\dot{x}(\bar u)\rangle d\bar u \quad (\leq 0). \ee

Moreover\footnote{This last part will be needed only in the next section, in order to prove global hiperbolicity.}, if $-H(x,u)$ behaves subquadratically (inequality (\ref{subquadratic}) holds with
$p(u) < 2$) then the same conclusion holds for any $\epsilon >0$, i.e., fixed also any $\epsilon >0$ there exists $R_2'>0$ such that (\ref{aux3}) holds for any such a curve $\alpha(u)$.
\end{lemm}
\noindent {\bf Proof.}
Put $E_\alpha (u) = \langle \dot \alpha (u), \dot \alpha (u)\rangle =
\langle \dot x (u), \dot x (u)\rangle +2\dot v (u)+ H(x,u)$, then:
\begin{equation}\label{ve1}
v(u)-v(u_0)=\frac{1}{2}\int_{u_{0}}^{u}\left( E_{\alpha}(\bar u)-\langle \dot{x}(\bar u),\dot{x}(\bar u)\rangle- H(x(\bar u),\bar u)\right) d\bar u, \quad \forall u\in [u_{0},u_1].
\end{equation}
On the other hand, using
the quadratic condition (\ref{subquadratic}):
\be \label{laleche}
-\int_{u_{0}}^{u}H(x(\bar{u}),\bar{u}) d\bar{u} \leq
R_1^{max} \int_{u_{0}}^{u}
d^{2}(x(\bar{u}), \bar x )d\bar{u} + R_2^{max} (u-u_0),
\ee
where $R_i^{max}$ is the maximum of $R_i(u)$ in $[u_0,u_0+\epsilon]$. Thus, choosing
$\epsilon>0$ small enough and taking into account that $x_0$ belongs to the bounded set $B$:
\begin{equation}\label{camp}
-\int_{u_{0}}^{u}(H(x(\bar{u}),\bar{u})+R'_{2})d\bar{u}<
\frac{1}{2\epsilon^{2}}\int_{u_{0}}^{u}
d^{2}(x(\bar{u}),x_{0})d\bar{u}\leq
\frac{1}{2}\int_{u_{0}}^{u}\langle\dot{x}(\bar{u}),\dot{x}(\bar{u})\rangle
d\bar{u},
\end{equation}
(the last inequality by (\ref{toto})) for some $R_2'>0$ and all $u\in [u_{0},u_{1}]$.
Therefore, we obtain
\begin{equation}\label{aux2}
-\int_{u_{0}}^{u}\left( \frac{1}{2}\langle
\dot{x}(\bar{u}),\dot{x}(\bar{u})\rangle
+H(x(\bar{u}),\bar{u})\right)d\bar{u}< R'_{2}
(u-u_{0}),\quad\forall u\in [u_{0},u_{1}].
\end{equation}
Thus, the result follows by using (\ref{aux2}) and (\ref{ve1}),
taking into account that $E_\alpha (u)\leq 0$.

Finally, in the subquadratic case, notice that fixed $\epsilon>0$,
(\ref{laleche}) holds replacing $d^{2}(x(\bar{u}), \bar x )$ by $d^{p}(x(\bar{u}), \bar x )$ with $p= $Max$\{p(u): u\in [u_0,u_0+\epsilon ]\} <2$. Thus, choosing $R_2'$ big enough, inequality (\ref{camp}) still holds.

\cvd
\smallskip

\begin{rema}\label{cojo} The constants $\epsilon, R'_{2}>0$ in Lemma \ref{lsc2} (or only $R'_2$ for its last assertion) can be chosen such that, fixed $u_{0}\in \R$, for any causal curve $\alpha(u) = (x(u), v(u), u)$, $u \in [u_0, u_1]$ with
$u_1-u_0 \leq \epsilon$ and $x(u_1)=x_1$ in a bounded subset $B_1$ of
${\cal M}_{0}$ (instead of $x(u_{0})=x_0 \in B$), one also has: \be
\label{aux33} v(u_{1}) -v(u) - R_2'(u_{1}-u) < -\frac{1}{4}
\int_{u}^{u_{1}} \langle \dot{x}(\bar u),\dot{x}(\bar u)\rangle
d\bar u \quad (\leq 0). \ee
\end{rema}

\noindent {\bf Proof of Theorem \ref{tsc}.} Fixed $z_0=
(x_{0},v_{0},u_{0})\in {\cal M}$ we can consider the basis of
neighborhoods of $z_0$, $\{U_{\epsilon'}: \epsilon' >0\}$, where
$U_{\epsilon'}=B(x_{0},\epsilon')\times
(v_{0}-\epsilon',v_{0}+\epsilon') \times
(u_{0}-\epsilon',u_{0}+\epsilon')$, and $B(x_0, \epsilon')$ is the
Riemannian metric ball  in $\mo$ centered at $x_0$ with radius
$\epsilon'$. Our aim is to prove that, fixed $\epsilon' >0$, there
exists $\epsilon \in (0,\epsilon']$ such that any causal curve
$\alpha(s) =(x(s),v(s),u(s))$ with $\alpha(0)=(x_{0},v_{0},u_{0})$
and endpoint $z_1= (x_{1},v_{1},u_{1}) \in U_{\epsilon}$ , lies
entirely in $U_{\epsilon'}$. Without loss of generality, one can
assume that $\alpha(s)$ is timelike with $\dot u(s) >0$ (the proof
if $\dot u(s)<0$ is analogous), and use $u$ to parametrize
$\alpha$, i.e., we will consider $\alpha(u) = (x(u), v(u), u)$ as
in
Lemma \ref{lsc2}, with 
$ u_1-u_0 <\epsilon \leq \epsilon'$.

Now, fixed $\epsilon'>0$, let us see that  some small $\epsilon>0$
can be chosen such that $x(u)$ lies in $B(x_0, \epsilon')$
\footnote{We remark that the  inequalities in the remainder of the
proof are needed only for some small $\epsilon$, thus, the first
conclusion of Lemma \ref{lsc2} will be claimed.}. When $u=u_1$,
the left-hand side of (\ref{aux3}) is equal to $(v_1-v_0)-
R_2'(u_1-u_0)$, thus, (\ref{aux3}) yields
\[
\int_{u_0}^u \langle \dot{x}(\bar u),\dot{x}(\bar u)\rangle d\bar
u < 4 (1+ R_2') \epsilon,
\]
for $\epsilon$ small enough. This inequality is a bound for the
energy and, thus, the length of $x(u)$, as required.

Finally, let us show that $v(s)$ remains in $(v_0-\epsilon', v_0+
\epsilon')$ for $\epsilon$ small enough. Notice that, from
(\ref{aux3}): \be \label{aux5a} v(u) -v_0 < R_2' (u_1-u_0) < R_2'
\epsilon. \ee Analogously, from (\ref{aux33}) \be \label{aux5b}
v_1 -v(u) < R_2' \epsilon, \ee and the bound for $v(u)$ follows
from (\ref{aux5a}), (\ref{aux5b}). As consequence, the causal
curve $\alpha (s)$ lies entirely in $U_{\epsilon'}$ and, thus, the
PFW is strongly causal. \cvd


\section{Sufficient hypotheses for Global Hyperbolicity} \label{s4}

This section is devoted to prove the following theorem, which completes our study of causality of PFW's.

\begin{theo} \label{tgh}
Any PFW 
with $\mo$ complete and $-H(x,u)$ subquadratic (according to (\ref{subquadratic})), is globally hyperbolic.
\end{theo}
The proof will be straightforward from the following two lemmas. The first one is an obvious  consequence of \cite[p. 409, Lemma 14]{O} (alternatively, see the proof of \cite[Corollary 3.32]{BEE}).

\begin{lemm} \label{lgh1}
Let $(M,g)$ be a strongly causal spacetime. If $J(p,q):=J^{+}(p)\cap J^{-}(q)$ is included in a compact subset $K \subset M$ then $J(p,q)$ is closed (and thus, compact).
\end{lemm}
The second one is valid if (\ref{subquadratic}) holds.

\begin{lemm} \label{lgh2}
If $-H$ behaves subquadratically  then the natural projections of $J(z_0,z_1)\subset {\cal M}_{0}\times \R^{2}$, $z_0<z_1$, on $\mo$ and $\R^2$ are bounded, for the distance $d$ associated to $\langle\cdot,\cdot\rangle$ on ${\cal M}_{0}$ and the usual distance $du^2 + dv^2$ on $\R^{2}$, respectively.
\end{lemm}
{\bf Proof.} As $J(z_0,z_1)\subseteq$ closure$(I(z_0,z_1))$, it is sufficient to prove the result for $I(z_0,z_1) (:=I^{+}(z_0)\cap I^{-}(z_1))$. Consider a point $r\in I(z_0,z_1)$.
From
(\ref{udot}), clearly $u(z_0)\leq u(r) 
\leq u(z_1)$ and, therefore, the points in $I(z_0,z_1)$ have component $u$ bounded.

In order to bound the component $v$, consider any future-directed
timelike curve $\alpha$ joining $z_0$ and $r$ (resp., $r$ and
$z_1$), and use the last assertion of Lemma \ref{lsc2}, plus Remark \ref{cojo}, to obtain\footnote{Notice that this is like (\ref{aux5a}), (\ref{aux5b}), but in that case we needed only inequalities for small $\epsilon$, and now $\epsilon$ must be bigger than $u(r)-u(z_0)$ and $ u(z_1)-u(r)$.}:
\[
v(r)-v(z_0)< R'_{2}(u(r)-u(z_0)),
\quad \quad v(z_1)-v(r)< R'_{2}(u(z_1)-u(r)),
\]
and $v(r)$ is also bounded. Finally, the bound of the projection of $r$ on $\mo$ follows by using again (\ref{aux3}), (\ref{aux33}) and the boundedness of $v(u), u$.
 \cvd
\smallskip

\noindent {\it Proof of Theorem \ref{tgh}}. From Theorem \ref{tsc}, we have just to prove that
$J(z_0,z_1)$ is compact for any $z_0<z_1$. The  Riemannian metric $g_R= \langle \cdot, \cdot \rangle + dv^2+du^2$ on $\mo \times \R^2$ is complete because of the completeness of the Riemannian distance $d$ on $\mo$. Thus, Lemma \ref{lgh2} implies that each $J(z_0,z_1)$ is included in a compact subset $K$ and, by Lemma
\ref{lgh1}, $J(z_0,z_1)$ is compact, as required.
\cvd

\begin{rema} {\em A well--known result by Avez and Seifert asserts that, in any globally hyperbolic spacetime, any pair of causally related points can be joined by means of a causal geodesic of maximum length. This property (in addition to multiplicity and other results) was obtained directly in \cite{CFSa}  by using the variational results in \cite{CFSb}. Assuming that it holds, the proof of Lemma \ref{lgh1} could be simplified by using that each point in
$J(p,q)$ can be connected to $p$ and $q$ by a causal geodesic.
}\end{rema}
As we have seen, there are counterexamples to global hyperbolicity in the quadratic case (classical plane waves) as well as in the superquadratic one (Proposition \ref{remnondis}). Nevertheless, previous proof can be used to study the possible global hiperbolicity of other quadratic or superquadratic PFW's. In fact, it is not difficult to construct globally hyperbolic examples
in these cases.
\begin{ex}\label{exglobhyp} {\rm
Consider any PFW with ${\cal M}_{0}=\R$ and $H(x,u)=H(x)$ satisfying, for some $p>0$ the
conditions:
\begin{itemize}
\item[(i)] $H(x)$ is constantly equal to $8(|n|+1)^{4p}$, on each interval $I_{n}=[n+1/4,n+3/4]$, $n\in\Z$, \item[(ii)]
$H(n)=-|n|^{p}$ is the minimum value of $H$ on each interval
$J_{n}=[n-1/4,n+1/4]$, $n\in\Z$.
\end{itemize}
Recall that, because of (ii), if $p = 2$ then $-H(x,u)$ behaves quadratically, and if $p>2$ superquadratically. Nevertheless, even in this case the PFW is globally hyperbolic. To check it,  a proof plainly analogous to the one of Theorem \ref{tgh} can be carried out,
say: both conclusions of Lemma \ref{lsc2}
(the one for small $\epsilon$ and the stronger for arbitrary $\epsilon$),
 plus Remark \ref{cojo}, still holds (in fact, one would
check that inequality (\ref{aux2}) holds by using the properties (i) and (ii) of $H$). Thus, as a consequence of the first conclusion one obtains strong causality, as in the proof of
Theorem \ref{tsc}, and the second conclusion yields global hiperbolicity, as in the proof of Theorem \ref{tgh}.
}\end{ex}

\section{Energy conditions and stress-energy tensor} \label{s45}

In order to give general geometrical results, no assumption on the
stress-energy tensor of PFW's has been done up to now. We will
study now the interplay between our asymptotic geometrical
conditions for $-H$ and reasonable matter sources. To this end,
first we will characterize when a PFW satisfy the classical energy
conditions. From this characterization (Proposition \ref{45a}), it
will be clear that these conditions are compatible with both,
subquadratic and superquadratic growth of -$H$ at spatial
infinity, and a pair of explicit examples are given (Example
\ref{ejemplo}). Then, a further discussion on the stress-energy
tensor is carried out. We argue that, for complete ${\cal M}_0$,
only subquadratic functions $-H$ should be taken into account, in
order to consider waves with negligible effects at infinity.

We will assume that Einstein's equation with zero cosmological constant
\begin{equation} \label{eei}
G:= {\rm Ric} -\frac{1}{2} S g = 8\pi T
\end{equation}
is satisfied. Recall that the Ricci tensor of a PFW is (see \cite{CFSa}):
\begin{equation}\label{rtr}
{\rm Ric} = \sum_{i,j=1}^n R^{(R)}_{i j} d x^i \otimes d x^j
-\frac{1}{2}\Delta_{x}H \, du \otimes du,
\end{equation}
where ${\rm R}^{(R)}_{ij}$ denote the components of the Ricci
tensor ${\rm Ric}^{(R)}$ of $(\mo, \langle\cdot,\cdot\rangle)$ in
the coordinates $(x^1,\dots , x^n)$. Notice also from (\ref{eei}), (\ref{rtr})
that the strong energy condition (or, equivalently,
the timelike convergence condition Ric$(\xi, \xi)\geq 0$ for all
timelike $\xi$) holds if and only if: \be \label{cct} {\rm
Ric}^{(R)}(\xi_0,\xi_0) \geq 0 , \; \forall \xi_0\in T{\cal M}_0,
\quad \quad \Delta_xH \leq 0. \ee

\begin{propo} \label{45a}
Let ${\cal M} = {\cal M}_0\times \R^2$ be a 4-dimensional PFW, and
let $K(x)$ be the curvature of ${\cal M}_0$. The following
conditions are equivalent.

(A) The strong energy condition (Ric$(\xi, \xi)\geq 0$ for all timelike $\xi$).

(B) The weak energy condition ($T(\xi, \xi)\geq 0$ for all timelike $\xi$).

(C) The dominant energy condition ($-T_b^a \xi^b $ is either 0 or
causal and future-pointing, for all future-pointing timelike $\xi\equiv \xi^b$).

(D) Both inequalities:
$$K(x) \geq 0 , \quad \Delta_xH(x,u) \leq 0, \quad \quad  \forall (x,u) \in {\cal M}_0\times \R.$$

\end{propo}
{\bf Proof.} As dim${\cal M}_0=2$, we have the following relations
for Ric$^{(R)}$ and the corresponding scalar curvature
$S^{(R)}$:
\be \label{ricesc} 
{\rm Ric}^{(R)}(\xi_0, \xi_0)= K(x)\cdot \langle
\xi_{0},\xi_{0}\rangle, \; \forall \xi_0 \in T_x{\cal M}_0 , \quad
S^{(R)}(x) = 2 K(x) , \; \forall x \in {\cal M}_0 .\ee

(A)$\Leftrightarrow$(D). The equivalence is clear from (\ref{cct}) and (\ref{ricesc}).

(B)$\Leftrightarrow$(D). Notice that $S^{(R)}$ and $S$ coincide
because of (\ref{rtr}); thus, Einstein tensor $G$ on any tangent
vector $\xi = (\xi_0, \xi_v, \xi_u) \in T_{(x,v,u)}{\cal M}$
becomes: \be \label{ggg} G(\xi,\xi) = K(x) \left(\langle \xi_0
,\xi_0 \rangle - \langle \xi,\xi \rangle\right) -
\frac{1}{2}\Delta_x H(x,u) \xi_u^2. \ee As the term in parentheses
is positive for timelike $\xi$, we have (D) $\Rightarrow $ (B).
For the converse, just notice that this term satisfies
$$ \langle \xi_0 ,\xi_0 \rangle - \langle \xi,\xi \rangle= -2 \xi_u \xi_v - H \xi_u^2 , $$
and $\xi_v$ can be chosen to make this (positive) term both, arbitrarily big and arbitrarily
close to 0.

(C)$\Leftrightarrow$(D).
Recall that the dominant energy condition (C) always implies the weak energy condition (B); even  more,  as $\xi^b$ is timelike and future-pointing,
when  $-T_b^a\xi^b $ is causal then it will be future-pointing if and only if
$0>-T_{ab}\xi^b \xi^a = -T(\xi,\xi)$. Thus, to check (D) $\Rightarrow $ (C), we only have to prove that, for any timelike $\xi$, (D) implies
\be \label{dominant}
g^{ac} G_{ab} \xi^b G_{cd} \xi^d \leq 0.
\ee
A straightforward computation from (\ref{ggg}) (or, equivalently, (\ref{e8pt})) shows:
that the left hand side of (\ref{dominant}) becomes:
$$ 2 K(x)^2 \xi_u \xi_v + \left(K(x) \Delta_xH(x,u) + K^2(x) H(x,u)\right) \xi_u^2 $$
$$= K(x)^2 \left(\langle\xi, \xi \rangle - \langle \xi_0,\xi_0 \rangle \right) + K(x) \Delta_xH(x,u) \xi_u^2, $$
which is non-positive because of (D).
\cvd

\smallskip
\begin{ex} \label{ejemplo} {\rm
There are well-known complete Riemannian surfaces with $K>0$, as the paraboloid. For simplicity, we will assume in the following examples ${\cal M}_0 = \R^2$, with its usual Riemannian metric.

(1) Assume that $H(x_1,x_2,u)$ is radial in $(x_1,x_2)$ or, equivalently, $H(x_1,x_2,u)$ is independent of $x_2$ ($x\equiv x_1$ can be then interpreted as a radial coordinate). Writting $H$ as $H_u(x)$, the energy conditions will hold if and only if
$$ \frac{d^2H_u}{dx^2} \leq 0 .$$
If the more restrictive condition
$$ \frac{d^2H_u}{dx^2} \leq -\epsilon < 0 $$
holds, then $-H_u(x)$ will be either quadratic or superquadratic at spatial infinity. If, say,
$$ -\frac{A(u)}{|x|^{q(u)}} \leq \frac{d^2H_u}{dx^2}(x) \leq 0 $$
for some $A(u), q(u) >0$ then it will be subquadratic (see Proposition \ref{45b} for a more general result).

(2) Let $H(x_1,x_2,u) = \alpha_u(x_1^2-x_2^2) f(u) + 2\beta_u(x_1
\cdot x_2) g(u)$ for some real  functions  $\alpha_u, \beta_u, f,
g$. Recall that when $\alpha_u \equiv \beta_u =$ Identity, this is
the classical example of gravitational plane wave. From a simple
computation, if $f, g\geq 0$ and $\alpha_u$ and  $\beta_u$ are
concave (i.e., $\alpha_u''\leq 0,  \beta_u''\leq 0$) then
$\Delta_xH \leq 0$, i.e., the energy conditions hold. Moreover,
if, for large $|s|$, $\alpha_u(s) = A(u) s^{1-q(u)} , \beta_u(s)=
\tilde A(u) s^{1-\tilde q(u)}$ and $0< q(u), \tilde q(u) < 1$ then
$-H$ is subquadratic (in principle, this may be a good subquadratic approximation to a plane wave, but notice Remark \ref{recont}). }\end{ex}

\noindent From (\ref{ggg}), the stress-energy tensor $T$ can be
written as
\begin{equation} \label{e8pt}
8\pi T = -K(x) (du\otimes dv + dv\otimes du) -(K(x)H(x,u) + \frac{1}{2}\Delta_xH(x,u)) du^2,
\end{equation}
thus, the PFW is vacuum if and only if $K\equiv 0, \Delta_xH \equiv 0$.
At the points with $\Delta_xH(x,u) = 0$, $T_a^b$ is diagonalizable, being the energy density $\rho$ and principal pressures $p_i$:
$$ \rho(x,v,u)  = \frac{1}{8\pi}K(x), \quad p_1(x,v,u) = - \frac{1}{8\pi} K(x), \quad p_2(x,v,u)= p_3(x,v,u) = 0. $$
Nevertheless, when $\Delta_xH(x,u) \neq 0$ then $T_a^b$ is {\em
not} diagonalizable, and it admits as a single eigenvalue $-K(x)$
(apart from 0 as a trivial double eigenvalue). But this does not seem to
be unreasonable when one takes into
account that the wave must have a finite extension or, say, its
effects must decrease towards infinity. In fact, in the  exact
model (\ref{quadratic}) the equality $\Delta_xH(x,u) = 0$ is due
to the fact that, at each $u$, the eigenvalues of Hess$_xH$ are
always equal and opposed. But these eigenvalues are constant in
the variable $x\in \R^2$ and, thus, the effects of the curvature at
infinite are not negligible. Recall that, for any unit $\xi_0 \in
T_x{\cal M}_0$, the sectional curvature of the plane $\Pi
(=\Pi(x,u,v)) = {\rm Span}(\xi_0,
\partial_u)$ (if non-degenerate, $H(x,u)\neq 0$) is: \be \label{ksks} K_S(\Pi) =
-\frac{1}{2H(x,u)}{\rm Hess}_xH[\xi_0,\xi_0] \ee (see
\cite[Section 2]{CFSa} for explicit computations of Christoffel
symbols) and, taking an orthonormal basis $(\xi_1, \xi_2)$ of
$T_x{\cal M}_0$ with the corresponding planes $\Pi_i = {\rm
Span}(\xi_i, \partial_u), $:
$$ {\rm Ric}(\partial_u,\partial_u) =  H(x,u) (K_S(\Pi_1)+ K_S(\Pi_2)). $$
In conclusion, it seems more realistic if, at each $u$, these sectional curvatures
(or even better, the eigenvalues of ${\rm Hess}_xH$) go to zero reasonably fast
when $|x|\rightarrow \infty$.
But in this case $-H$ should be subquadratic. More precisely:

\begin{propo} \label{45b}
Let ${\cal M} = {\cal M}_0\times \R^2$
be a 4-dimensional PFW with complete
${\cal M}_0$.
For each $u$, let $\lambda_i(x), i=1,2$, be the eigenvalues of Hess$_xH(\cdot , u)$
and $\lambda(x) ={\rm Min}\{\lambda_1,(x), \lambda_2(x)\}$. Assume that
\be \label{eal} -\frac{A}{d(x,\bar x)^q} \leq \lambda(x) \ee
for some $A, q>0$ (which may depend on $u$) and $\bar x \in {\cal M}_0$.

Then, $-H(x,u)$ satisfies the subquadratic relation (\ref{subquadratic}).
\end{propo}
{\bf Proof}. Let $H_0(u) = {\rm Min}\{H(x,u) : \; d(x,\bar x) \leq 1\}$ and,
for each $x$, consider a minimizing unit geodesic $\gamma_x$ from $\bar x$ to $x$. Without loss of generality, we can assume $0<q<1$ and, for any $x$ with
$d(x,\bar x) > 1$,
$$-(H(x,u)  -H_0(u)) \leq
- \int_1^{d(x,\bar x)} \int_1^{\bar s} \frac{d^2}{ds^2} H(\gamma_x(s),u)ds d\bar s
$$
$$
= -\int_1^{d(x,\bar x)} \int_1^{\bar s} {\rm Hess}_xH[\gamma'_x(s),\gamma'_x(s)]ds d\bar s
$$
$$
\leq -\int_1^{d(x,\bar x)} \int_1^{\bar s} \lambda(\gamma_x(s))ds d\bar s
\leq A \int_1^{d(x,\bar x)} \int_1^{\bar s} \frac{1}{d(\gamma_x(s),\bar x)^q}ds d\bar s $$
$$
= A \int_1^{d(x,\bar x)} \int_1^{\bar s} \frac{1}{s^q}ds d\bar s
< \frac{A}{(1-q)(2-q)} d(x,\bar x)^{(2-q)} + \frac{A}{(1-q)}
$$
which is subquadratic, as required. \cvd
\smallskip

\begin{rema} \label{recont} {\rm
Summing up, for complete ${\cal M}_0$:
(a) the inequality for the eigenvalues of Hess$_xH$, $\lambda_1 + \lambda_2 \leq 0$ is equivalent to $\Delta_xH \leq 0$, (b) in this case, $K\geq 0$ is equivalent to the energy conditions and (c) independently, if inequality (\ref{eal})
(which is implied by a natural sense of assymptotic flatness)
holds then $-H$ is subquadratic and the PFW globally hyperbolic.
Then, these three items are physically reasonable, and they are
satisfied by simple examples, as those in Example \ref{ejemplo}(1). Nevertheless, the third item is not satisfied by Example \ref{ejemplo}(2) (even though $\alpha_u, \beta_u$ may have a good behaviour at infinity, when, say, $x_1=x_2$ the eigenvalues of Hess$_xH$ for large $x_i$ depends on the second derivative of $\alpha_u(s)$ at 0). This should be taken into account for realistic models of plane waves at infinity.}
\end{rema}

\section{Conjugate points} \label{s5}

In order to obtain a detailed study on existence and multiplicity of connecting geodesics in exact gravitational waves, Ehrlich and Emch introduced in \cite{EE1} the concept of ``first astigmatic conjugate pairs'' for the coordinate $u$ (defined for an ODE system  in \cite[Definition 3.12]{BEE}). In that reference they concluded: (a) the existence of a unique connecting geodesic (which is causal for causally related points) whenever  $u_1$ appears before the first astigmatic conjugate point of $u_0$, and (b) the non-geodesic connectedness when $u_1$ is the first astigmatic conjugate point. Moreover, in the case (b) the  points in $\Pi_{u_{1}}=\{ (x,v,u_1): x\in \mo, v \in \R\}$ reached by geodesics are conjugate to the initial point $z_{0}$.
The conclusions in \cite[Subsection 4.3]{CFSa} extend and complement these results, yielding in particular a explicit bound for the appearance of the first conjugate pair.

In this section we will see how, in general PFW's, one can still speak on conjugate pairs $(x_0,u_0), (x_1,u_1)$ for a suitable trajectory joining $x_0$ and $x_1$, Definition \ref{deff}. This notion is related to the usual conjugate points of  geodesics (Proposition \ref{reeff}), and we also discuss  how  yields the notion of conjugate (eventually astigmatic) pair $(u_0, u_1)$, in the particular case of a classical gravitational plane wave. Moreover, we explain how the existence of conjugate points can be studied systematically as a ``purely Riemannian'' problem (Proposition \ref{prorfin}, Remark \ref{remfin}).

\smallskip

\noindent Recall first that, from geodesic equations, a curve
 $z:\ ]a,b[\ \rightarrow \m$,
$z(s)= (x(s), v(s), u(s))$ ($\ ]a,b[\ \subseteq \R$),
 with constant $\langle \dot z(s), \dot z(s)\rangle = E_z$
is a geodesic  if and only if the three following conditions hold:
(a) $u(s)$ is affine, i.e.,
$u(s) = u_0 + s \Delta u$ for some
$u_0, \Delta u  \in \R$, $\Delta u \equiv \dot u(s)$,
(b) $x = x(s)$ is a solution of:
\begin{equation}
\label{RiemannianEq}
D_s\dot x = - \nabla_x V_{\Delta}(x(s),s) ,
\end{equation}
where
\begin{equation} \label{eV}
V_{\Delta}(x,s) = -\ \frac{(\Delta u)^2}{2}\ H(x, u_0 + s \Delta u);
\end{equation}
and (c) when  $\Delta u = 0$, $v = v(s)$ is affine,
otherwise:
\begin{equation} \label{ev}
v(s) = v_0 + \frac{1}{2 \Delta u} \int_0^s \left( E_z - \langle \dot x(\sigma), \dot x(\sigma)\rangle +
2 V_{\Delta}(x(\sigma), \sigma)\right) d\sigma.
\end{equation}
(see \cite{CFSa} for details). Thus, the component $x(s)$ can be seen as a critical point of a functional as follows.
Fixed two points $x_0$, $x_1 \in \mo$, define the set
$\Omega^1(x_0,x_1) $ containing the curves $x(s)$,  $x: [0,1] \rightarrow \mo$, $x(0) = x_0, x(1) = x_1$,
(for consistency, it is convenient to assume that each curve $x(s)$ is just
 absolutely continuous with finite length).
From
(\ref{RiemannianEq}), the projections $x(s)$ of  geodesics $z:[0,1] \rightarrow \m$
with fixed $z(0)=(x_0,v_0,u_0)$, $z(1)=(x_1,v_1,u_1), u_1=u_0+\Delta u$, are in bijective correspondence with the
critical points of the functional ${\cal J}_{\Delta }$,
\begin{equation} \label{eJ}
{\cal J}_{\Delta}: x \in \Omega^1(x_0,x_1) \longmapsto
\ {1\over 2}\ \int_0^1 \langle\dot x(s),\dot x(s)\rangle\ ds
\ - \int_0^1 V_{\Delta}(x(s),s)\ ds \in \R,
\end{equation}
where $V_{\Delta}$ is defined in (\ref{eV}). Notice that the role of $v_0, v_1$ is irrelevant for ${\cal J}_{\Delta}$, while $u_0, u_1$ play a role through the expression of $V_{\Delta}$.

Now, we can state the following definition, in agreement with the second variation for critical values of functionals (see for example \cite[Chapters 4,5]{GH}):
\begin{defi}\label{deff} Fix $\overline{z}_{0}=(x_{0},u_{0})$, $\overline{z}_{1}=(x_{1},u_{1}) \in \m_{0} \times \R$, and let $x(s)$ be a critical point of ${\cal J}_{\Delta}$ with endpoints $x_{0},x_{1}$ and $\Delta u = u_1-u_0$. We say that  $\overline{z}_{0}$, $\overline{z}_{1}$ are conjugate points along $x(s)$ of multiplicity $\overline{m}$ if the dimension of the nullity of the Hessian of ${\cal J}_{\Delta}$ in $x(s)$ is $\overline{m}$ (if $\overline{m}=0$ we say that $\overline{z}_{0}$, $\overline{z}_{1}$ are not conjugate).
\end{defi}
Of course, the usual interpretation for conjugate points of
geodesics holds for this definition, i.e., essentially, if $\bar
z_0, \bar z_1$ are conjugate along $x(s)$ for ${\cal J}_{\Delta}$
then they are ``almost meeting points'' for the solutions of
(\ref{RiemannianEq}) in some direction, being the multiplicity
equal to the number of such independent directions. In fact, by a
direct computation, the Jacobi equation for trajectories under the
potential $V_{\Delta}$ under a variation of $x(s)$ is:
\begin{equation}\label{sisi}
\frac{D^{2}\overline{J}}{d s^{2}}+R_0(\dot x(s),\overline{J})\dot x(s)+Hess_{x}V_{\Delta}(\overline{J},\cdot, s)^{\flat}=0,
\end{equation}
where $R_0$ denotes the curvature of $\mo$ (which is equal to the restriction of the curvature $R$ on $\m$) and $^{\flat}$ denotes the vector field on ${\cal M}_{0}$ metrically associated to the corresponding $1$-form (that is, $\langle w,Hess_{x}V_{\Delta}(\overline{J},\cdot,s)^{\flat}\rangle=Hess_{x}V_{\Delta}(\overline{J},w,s)$, for all $w\in T\mo, s\in \R$). A standard computation shows that $\overline{z}_{0}$, $\overline{z}_{1}$ are conjugate of multiplicity $\overline{m}$ according to Definition \ref{deff} if and only if  the dimension of the Jacobi fields satisfying (\ref{sisi}) with $\overline{J}(0)=\overline{J}(1)=0$ is $\overline{m}$.
Such a Jacobi field $\overline{J}$ is the variational field of a variation $x_t(s)$ of $x(s)$ through curves $s\rightarrow x_t(s)$ which satisfy (\ref{RiemannianEq}), i.e., $\overline{J}(s) =
\partial_t|_0 x_t(s)$, where $x_0(s)=x(s)$ for all $s$ (notice that, even though $\overline{J}(1)=0$,
the value of $x_t(1)$ might be different to $x(1)$ and, thus, $x_t$ not necessarily belongs to
$\Omega^1(x_0,x_1)$).
Additionally, recall that $z_{0},z_{1}\in\m$ are conjugate of multiplicity $m$ along a geodesic $z(s)$ if and only if the dimension of the Jacobi fields $J(s)$ satisfying
\begin{equation} \label{ejac}
\frac{D^{2}J}{d s^{2}}+R(\dot z(s),J)\dot z(s)=0,
\end{equation}
with $J(0)=J(1)=0$ is $m$. Essentially, the projection of the independent directions
of conjugacy of $z(s)$ as a geodesic are the independent directions
of conjugacy of $x(s)$ according to Definition \ref{deff}:

\begin{propo}\label{reeff} The pairs $\overline{z}_{0}=(x_{0},u_{0})$, $\overline{z}_{1}=(x_{1},u_{1})$ are conjugate of multiplicity $\overline{m}$ along $x(s)$ (according to Definition \ref{deff}), if and only if for any  geodesic  $z: [0,1]\rightarrow \m$ with $z(s)=(x(s),v(s),\Delta u\cdot s+u_{0})$ the corresponding endpoints $z_{0}=(x_{0},v_{0},u_{0})$, $z_{1}=(x_{1},v_{1},u_{1})$ are conjugate with the same multiplicity $m=\overline{m}$.
\end{propo}

\noindent {\bf Proof.} Let $V$ (resp. $\overline{V}$) be the $m$-dimensional (resp. $\overline{m}$-dimensional) space of the Jacobi fields on $z(s)$ (resp. $x(s)$) with $J(0)=J(1)=0$ (resp. $\overline{J}(0)=\overline{J}(1)=0$). We will denote by $\pi_{x}$ (resp. $\pi_{v}$, $\pi_{u}$) the usual projection of $T\m$ on $T\m_0$ (resp. $T\R_{v}\equiv \R$, $T\R_{u} \equiv \R$). Given $J\in V$, we will put $\hat J = \pi_x(J), J_u = \pi_u(J), J_v = \pi_v(J)$. As $\partial_v$ is parallel, from (\ref{ejac}) one has
\begin{equation}\label{ju}
 \pi_u \left(\frac{D^2J}{ds^2}\right) \equiv 0, \quad \quad \hbox{and} \quad \quad J_u  \equiv 0.
\end{equation}
In what follows, our aim is to prove $m=\overline{m}$, and we will assume $\Delta u \neq 0$ (otherwise, the result would be straightforward).

Fix $\overline{J} \in \overline{V}$ associated to a variation $x_t(s)$.
Now, consider the variation of $z(s)$ by geodesics $z_{t}(s)=(x_{t}(s),v_{t}(s),u_{0}+\Delta u\cdot s)$, where $v_{t}(s)$ is taken from (\ref{ev}) for $x(s)= x_t(s)$ and some $E_z = E_z^t$. Choosing:
\[
E_{z}^{t}=2\cdot \Delta u(v_{1}-v_{0})+\int_{0}^{1}(\langle \dot{x}_{t}(\sigma),\dot{x}_{t}(\sigma)\rangle -2V_{\Delta}(x_{t}(\sigma),\sigma))d\sigma,
\]
then $v_{t}(1)=v_{1}$ for all $t$. Therefore, the corresponding variational field $J$ belongs to $ V$ and $\hat J = \overline{J}$. Thus, $\overline{m} \leq m$.

In order to prove the reversed inequality, we have to check that any $J\in V$ can be reconstructed  from $\hat J \in \overline{V}$ as above. Otherwise, taking into account (\ref{ju}), we could find $J\in V$ non identically null with $\pi_x(\frac{DJ}{ds})(0)=0$. As $\hat J$ satisfies
 (\ref{sisi}), then necessarily $\hat J \equiv 0$ and $J (\equiv J_v)$ is the variational field of $z_t(s)= (x(s),v_t(s),u_0+\Delta u \cdot s)$, for some $v_t(s)$ where each $v_t$ satisfies (\ref{ev}) with $E_z = E_z^t$. But, as $x(s)$ is a critical point of ${\cal J}_\Delta$ in (\ref{eJ}),
equation (\ref{ev}) implies $\partial_t|_0v_t(s) = \partial_t|_0 E_z^t \cdot s/2\Delta u$, for all $s$. As
$\partial_t|_0v_t(1)$ must vanish, we obtain $J \equiv 0$, as required.
\cvd
\smallskip

\begin{rema} \label{rnuevo} {\em  The component $v(s)$ of the geodesic $z(s)$ in Proposition \ref{reeff} satisfies (\ref{ev}) for some arbitrary value of $E_z$. Thus, its causal character can be chosen timelike, spacelike or lightlike.
}
\end{rema}

\noindent Let us discuss the case of a plane wave; in particular,
(\ref{planefrontedmetric}) holds and $R_0(\dot x(s),\overline{J})\dot x(s) \equiv 0$. Writing, as usual, $f= h_{11} = -h_{22}, g= h_{12} = h_{21}$ in (\ref{quadratic}),
$$
Hess_{x}V_{\Delta} = -(\Delta u)^2 \left(
\begin{array}{rr}
f & g \\
g & -f
\end{array}
\right)(u),
$$
which is independent of $x$. Thus,   equations (\ref{sisi}), (\ref{ejac})  depends only on $u$, and the multiplicity of conjugation along a geodesic $z(s)$ joining two points $(y_{0},w_{0},v_{0},u_{0})$, $(y_{1},w_{1},v_{1},u_{1})$ will be independent not only of  $v_{0}$, $v_{1}$ but also of  $y_{0}$, $w_{0}$, $y_{1}$, $w_{1}$ and the particular geodesic chosen. This property is due to the particular symmetries of classical plane waves, and allows one to define ``when  $u_{0}$, $u_{1}$ are conjugate pairs''. 
In fact, in terms of Definition \ref{deff}, we can say that $u_{0}$, $u_{1}$ are conjugate pairs if the corresponding points $\overline{z}_{0}=(y_{0},w_{0},u_{0})$, $\overline{z}_{1}=(y_{1},w_{1},u_{1})$, for some (and then for any) $y_{0}$, $w_{0}$, $y_{1}$, $w_{1}$ are conjugate points along some (and then any) critical point $x(s)$ of ${\cal J}_{\Delta}$. This definition has obvious consequences from Proposition \ref{reeff} and Remark \ref{rnuevo}.

On the other hand, Ehrlich and Emch \cite{EE3} also introduced a notion of astigmatic conjugacy (in opposition to the less generic anastigmatic case): $u_{0}$ and $u_{1}$ are astigmatic conjugate if and only if they are conjugate with multiplicity $1$ (see \cite[Definition 3.12]{BEE}). Even though astigmatic conjugacy of pairs $u_0, u_1$ is related to properties on existence and multiplicity of geodesics (and thus, to the focusing of lightlike geodesics),  it makes sense only in very particular cases --essentially just in classical plane waves.
Recall that, in a general semi--Riemannian manifold (Lorentzian, Riemannian or with any index), there is no relation between the existence of conjugate points and the existence or global uniqueness of possible connecting geodesics. This is also valid for general PFW's and, thus, these questions are independent of the existence of conjugate pairs $\overline{z}_0=(x_0, u_0), \overline{z}_1= (x_1, u_1)$.  In fact, recall the following trivial examples: (i) Take any complete Riemannian manifold $({\cal M}_{0},\langle\cdot,\cdot\rangle)$ with conjugate points along some geodesic; then  the corresponding PFW obtained by taking $H\equiv 0$ is always geodesically connected (and globally hyperbolic), even though there exist conjugate points. (ii) Take as  $({\cal M}_{0},\langle\cdot,\cdot\rangle)$ the usual $2$-dimensional cylinder; then the corresponding PFW obtained by taking $H\equiv 0$ does not present uniqueness of connecting geodesics, even though there are no conjugate points. Nevertheless, the ``local uniqueness'' of connecting trajectories in absence of conjugate points (i.e., if there are no conjugate points then there exists a neighborhood of the connecting geodesic where it is unique)  holds in any semi--Riemannian manifold. Thus, only the corresponding property related to ``astigmatic conjugate pairs'' in exact gravitational waves can be extended, by  Proposition \ref{reeff}, to conjugate pairs $(\overline{z}_0, \overline{z}_1)$ in arbitrary PFW's.

\smallskip

\noindent
For a general PFW, Proposition \ref{reeff} allows to study conjugate points as a purely Riemannian problem. Thus, it is possible to conclude the non-existence of conjugate points by imposing certain conditions on the sign of the sum of the sectional curvature of the planes plus the Hessian of $V_{\Delta}$. In fact, we wonder if there is a solution of (\ref{sisi}) with $\overline{J}(0)=0$ ($\overline{J}'(0)\neq 0$) and also vanishing at some $s_{0}>0$. Recall that
\[
\begin{array}{ll}\langle \overline{J},\overline{J}\rangle'' & =2\langle \overline{J}',\overline{J}'\rangle +2\langle \overline{J}'',\overline{J}\rangle \\ & =2\langle \overline{J}',\overline{J}'\rangle -2\langle R(\dot x(s),\overline{J})\dot x(s),\overline{J}\rangle - 2\langle Hess_{x}V_{\Delta}(\overline{J},\cdot)^{\flat},\overline{J}\rangle \\ & =2|\overline{J}'|^{2}-2K(\dot x(s),\overline{J})\cdot |\dot x(s)\wedge \overline{J}|^{2}-2Hess_{x}V_{\Delta}(\overline{J},\overline{J}),
\end{array}
\]
where $K(\dot x(s),\overline{J})$ is the sectional curvature of the plane spanned by $\dot x(s)$ and $\overline{J}$, and $|\dot x(s)\wedge \overline{J}|$ is the area of the parallelogram spanned by these same vectors. If
\begin{equation} \label{deshescur}
-2K(\dot x(s),\overline{J})\cdot |\dot x(s)\wedge \overline{J}|^{2}-2Hess_{x}V_{\Delta}(\overline{J},\overline{J})\geq 0
\end{equation}
for all $s$, then $\langle \overline{J},\overline{J}\rangle''\geq 0$ and
\be \label{ejj}
\langle \overline{J},\overline{J}\rangle '(s_{2})\geq \langle \overline{J},\overline{J}\rangle '(s_{1}) \quad \quad \hbox{whenever} \quad s_{2}>s_{1}.
\ee
Since $\langle \overline{J},\overline{J}\rangle'(0)=0$ and $\langle \overline{J},\overline{J}\rangle''(0)>0$ (recall $\overline{J}'(0)\neq 0$), it follows that, for any small positive  $s$,
\be \label{ejs0}
\langle \overline{J},\overline{J}\rangle (s)>\langle \overline{J},\overline{J}\rangle (0).
\ee
Therefore, from (\ref{ejj}) and (\ref{ejs0}),  $\langle \overline{J},\overline{J}\rangle(s)>0$ for all $s>0$, and $x(s)$ is not conjugate to $x(0)$ along $x(s)$. In particular,  taking into account that $V_\Delta$ in inequality (\ref{deshescur}) is essentially equal to $- H$:

\begin{propo} \label{prorfin} If $H$ is spatially convex (i.e. Hess$_xH(w,w,s)\geq 0$, $\forall w \in T\mo, s\in \R$) and the sectional curvature $K$ is non-positive, then no geodesic in a PFW has conjugate points.
\end{propo}
Notice that the spatial convexity of $H$ and the sign of $K$
points out the wrong direction with respect to the
energy conditions, which imply $\Delta_xH \leq 0$, $K\geq 0$. 
As expected, these conditions go in the direction to imply the existence of conjugate points.

\begin{rema} \label{remfin} {\em With more generality, the appearance of conjugate points for ${\cal J}_\Delta$ can be controlled by comparing with a model space, in the spirit of the comparison theorems for conjugate points of geodesics in Riemannian manifolds, which have as  starting point  Rauch's comparison theorem (see for example \cite{ChE} or \cite{Ch}). For example, it is not hard to prove the following result. Let ${\cal M}_{0}^{n}, \tilde{{\cal M}}_{0}^{n+k}$, $k\geq 0$ be Riemannian manifolds such that their sectional curvatures $K, \tilde K$ (resp.) satisfy $$K \leq 0 \leq \tilde K,$$ and let $V_\Delta, \tilde V_\Delta$ be potentials on ${\cal M}_{0}^{n}, \tilde{{\cal M}}_{0}^{n+k}$ (depending on time $t$) such that
$$Hess V_{\Delta}(w,w, t) \leq \mu(t) \leq Hess \tilde{V}_{\Delta}(\tilde{w},\tilde{w}, t),$$
for all unit $w, \tilde{w}$ and some function $\mu$. Let
$\gamma:[0,a]\rightarrow {\cal M}_{0}^{n}$ and $\tilde{\gamma}:[0,a]\rightarrow \tilde{{\cal M}}_{0}^{n+k}$, be critical points of the analogous corresponding functionals ${\cal J}_{\Delta}$, $\tilde{{\cal J}}_{\Delta}$, respectively, and let $\overline{J}$ and $\widetilde{\overline{J}}$ be Jacobi fields along $\gamma$ and $\tilde{\gamma}$ (resp.), such that
\[
\overline{J}(0)=\widetilde{\overline{J}}(0)=0,\;\; |\overline{J}'(0)|=|\widetilde{\overline{J}}'(0)|.
\]
If $\tilde{\gamma}$ does not have conjugate points on $(0,a]$
then
\[
|\widetilde{\overline{J}}(s)|\leq |\overline{J}(s)|.
\]
for all $s\in (0,a]$.

Such a comparison among Jacobi fields can be used to ensure the
existence (or inexistence) of conjugate points. In fact, if no
conjugate point appears for $\tilde{{\cal M}}_{0}^{n+k}$, then no
conjugate point on ${\cal M}_{0}^{n}$ will appear and, conversely,
if there are conjugate  points on ${\cal M}_{0}^{n}$ then this
forces the existence of conjugate points on $\tilde{{\cal
M}}_{0}^{n+k}$ (assuming that the critical curves are defined for
sufficiently big values of $s$). In fact, such a comparison result
with $\tilde {{\cal M}}_{0}^{n+k} \equiv \R^{n+k}$ and $\tilde
V_{\Delta} \equiv 0$ yields, as a particular case, Proposition
\ref{prorfin}.

Moreover, consider $\R^2 = {\cal M}_{0}^{n}= \tilde{{\cal M}}_{0}^{n+k}$ and compare:
(i) $V_{\Delta}$, constructed from $H$ for a plane wave $(\ref{quadratic})$, with
$f= h_{11} =-h_{22} \equiv \epsilon$, and (ii)
$\tilde V_{\Delta}$ constructed from some $\tilde H$ with eigenvalues $\tilde \lambda_i(x)$. If
\be \label{9999}
\tilde \lambda_i(x) \leq -\epsilon  \quad i=1,2
\ee
 (and, in particular, the energy conditions hold)
then there will be necessarily conjugate points for
$\tilde V_{\Delta}$, which will appear not later than those for $V_{\Delta}$. Notice that if (\ref{9999}) holds on all $\R^2$ then $-H$ will not be subquadratic. But, of course, one can modify $H$ after the appearance of conjugate points to obtain any desired asymptotic behavior, compatible with the energy conditions too.
} \end{rema}

Finally, it is worth pointing out that the Morse theory for geodesics in a PFW is then reduced essentially to Morse theory for Riemannian trajectories under potential $V_{\Delta}$, which is well-known \cite{BM}. 

\section{Conclusion}
Our results and conclusions in this paper can be summarized as follows.

In spite of the importance of the classical focusing property of null geodesics for the classical model of plane wave, this property depends strongly on idealizations such as the infinite extension of the wave and, thus, it must be regarded as unrealistic. The possible appearance of focalization is related to the existence of conjugate points, and this should be studied, in general, as the conjugate points for a classical Riemannian potential as in Section \ref{s5}.

The qualitative behaviour of causality and other properties of PFW's change dramatically depending on if $-H(x,u)$ behaves or not subquadratically at spatial infinity. In fact, we have the following possibilities:

(1) $-H$ subquadratic  and complete ${\cal M}_0$ $\Rightarrow$ globally hyperbolic.

(2) $-H$ quadratic   $\Rightarrow$ strong causality (there are
both, globally hyperbolic and non-globally hyperbolic examples).

(3) $-H$ superquadratic  $\Rightarrow$ causal (there are
examples  non-distinguishing as well as globally hyperbolic).
\smallskip

Therefore, when a PFW is taken as a model of a spacetime, one must
specify which asymptotic behaviour is expected to hold. In
principle, the subquadratic behaviour seems reasonable as a
consequence of asymptotic flatness and it is compatible with the
energy conditions; then, realistic PFW's should be regarded as
globally hyperbolic (or at least strongly causal, if an incomplete
$\mo$ were chosen). The implications of these results must be
taken into account in other important questions, as the
specification of Cauchy data or quantization.

\end{document}